# Dark defect charge dynamics in bulk chemical-vapor-deposition-grown diamonds probed via nitrogen vacancy centers

A. Lozovoi[1], D. Daw[1,2], H. Jayakumar[1], C. A. Meriles[1,2,*]

[1]Department of Physics, CUNY - The City College of New York, New York, NY 10031, USA.
[2]CUNY-Graduate Center, New York, NY 10016, USA.

**Abstract**

*Although chemical vapor deposition (CVD) is one of the preferred routes to synthetic diamond crystals, a full knowledge of the point defects produced during growth is still incomplete. Here we exploit the charge and spin properties of nitrogen-vacancy (NV) centers in type-1b CVD diamond to expose an optically and magnetically dark point defect, so far virtually unnoticed despite an abundance comparable to (if not greater than) that of substitutional nitrogen. Indirectly-detected photo-luminescence spectroscopy indicates a donor state 1.6 eV above the valence band, although the defect's microscopic structure and composition remain elusive. Our results may prove relevant to the growing set of applications that rely on CVD-grown single crystal diamond.*

The development of devices that exploit diamond for applications in emerging areas such as power electronics[1,2], quantum information processing[3-7], or nanoscale sensing[8-13] hinges on a detailed understanding of the material point defects and their interplay, particularly in the presence of injected or photo-generated charge carriers. This knowledge is especially important in samples created via chemical vapor deposition (CVD), one of the leading growth methods for applications that rely on synthetic single crystals. While various doping techniques have been developed to shape the properties of the host material (e.g., carrier mobility), pure CVD-grown diamond suffers from unwanted contamination during growth (e.g., due to impurities in the gas mixture or from incomplete reaction at the crystal surface), typically leading to the formation of nitrogen- or proton-based complexes[14-17]. Despite much progress[18-21], identifying and cataloguing these CVD-specific defects has proven to be a time-consuming, difficult task, particularly in the case of complexes that do not fluoresce or carry a net electronic spin.

Here we capitalize on the unique properties of the nitrogen-vacancy (NV) center[22] — a fairly common color center comprising a substitutional nitrogen immediately adjacent to a vacant site — to investigate the dynamics of trapped charge in bulk type 1b CVD-grown diamond. The combined use of fluorescence microscopy, optically detected multi-spin magnetic resonance, and photo-luminescence (PL) excitation spectroscopy points to a background defect — here referred to as X — whose concentration is at least comparable to substitutional nitrogen[23] (normally the most abundant in CVD diamond). We find that defect X has no emission in the visible or near-infrared and is (likely) spin-less, probably the reasons why it has remained relatively unnoticed despite its ubiquity. Though we presently cannot derive a concrete atomistic model, our results suggest this point defect may correspond to a proton complex with an acceptor level 1.6 eV above the valence band. Our results are consistent with a prior study[24] postulating the presence of defect X to regain balance in the transfer of charge between $N^0$ and $NVH^-$ in CDV type 1b diamond.

We conduct our experiments in a set of [100] type-1b diamond crystals (3×3×0.1 mm$^3$) grown via CVD with a nominal nitrogen concentration of ≲1 ppm and no post treatment. For consistency, we limit our description to the results attained with a CVD-grown sample from Delaware Diamond Knives, though we emphasize virtually identical responses were observed in other type-1b samples with comparable (nominal) nitrogen concentration produced by the same and other manufacturers (e.g., E6). In these samples, there is typically 1 NV every one hundred nitrogen impurities[25], i.e., the NV concentration is ≲ 10 ppb. To conduct our experiments, we use a custom-made multi-color optical microscope described before[26]. Briefly, this system integrates a 532 nm diode laser and a 632 nm He:Ne laser into a 0.43 numerical aperture objective; fluorescence detection is carried out via an avalanche photo-detector upon collection into an optical fiber. To separate the emission from neutral and negatively-charged NVs, we use a passband (550–620 nm) and a high-pass (>635 nm) filter, respectively. We conduct optically-detected magnetic resonance experiments (ODMR) with the help of 25-μm-diameter copper wire overlaid on the diamond surface and implement time-resolved measurements using a multi-purpose pulse generator.

To gather initial information on the defect composition of our sample, we first implement a double electron-electron resonance[27] (DEER) protocol with the NVs serving as a the optically-detected local spin probe (Fig. 1a). This strategy — already exploited in the past to study both bulk crystals and diamond nanoparticles[28-31] — uses green laser pulses to initialize and readout the NV spin state, as well as synchronous microwave manipulation of the NV$^-$ spin (MW1) and surrounding paramagnetic defects (MW2). Since the latter are a priori unknown, one can gather valuable information by monitoring the DEER signal — defined as the difference between the NV$^-$ spin echo response with and without the DEER pulse — as a function of the MW2 frequency.

Fig. 1b shows the results within a frequency window around that expected for spin-1/2 defects at the applied magnetic field

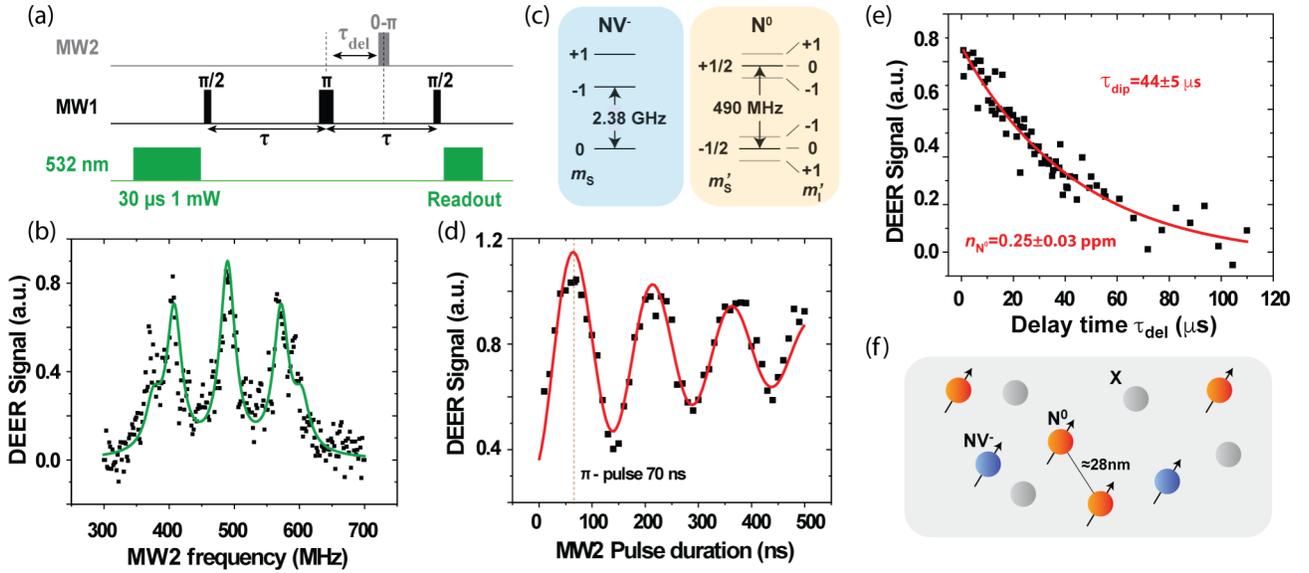

**Figure 1.** (a) DEER pulse sequence adapted to include a variable delay time $\tau_{del}$ between the MW1 and MW2 inversion pulses. (b) Measured DEER signal as a function of the MW2 frequency for $\tau = 57$ μs and $\tau_{del} = 0$ μs; the solid line is a calculated spectrum using the known $^{14}$N hyperfine coupling constants at the $N^0$ site. (c) Ground state energy diagram of the $NV^-$ and $N^0$ centers (left and right respectively) in a magnetic field of 17 mT aligned with the NV symmetry axis; $m_S$ ($m'_S$) denotes the $NV^-$ ($N^0$) electron spin projection number and $m'_I$ is the $^{14}$N spin projection number at the $N^0$ site. (d) DEER signal at 490 MHz as a function of the MW2 pulse duration; all other conditions as in (b). The solid trace is a damped sinusoidal and serves as a guide to the eye. (e) DEER signal (at the $m'_I = 0$ transition) as a function of $\tau_{del}$ for $\tau = 110$ μs; the solid line indicates an exponential fit with a time constant $\tau_{dip} = 44 \pm 5$ μs corresponding to an $N^0$ concentration $n_{N^0} = 0.25 \pm 0.03$ ppm. (f) $NV^-$ coexist with $N^0$ and a dark defect, referred to as X, at least as abundant as nitrogen, see below.

(17 mT), here oriented parallel with the NV symmetry axis. The observed spectral features — corresponding to $^{14}$N-induced hyperfine shifts — unambiguously point to the presence of $N^0$ centers (i.e., spin-1/2 paramagnetic impurities formed by neutral substitutional nitrogen, also known as 'P1', see Fig. 1c). Further, because the amplitude of the central peak relative to the satellites coincides with that expected for $N^0$ centers alone (see superimposed solid trace in Fig. 1b), these same observations also allow us to rule out other coexisting spin-1/2 impurities (which would otherwise make the central peak comparatively more prominent[29]).

To determine the actual nitrogen concentration, we first calibrate the duration of the DEER pulse (Fig. 1d) and subsequently vary the time interval $\tau_{del}$ between the π-pulses (Fig. 1e). The observed single-exponential response suggests that each NV center is coupled to multiple P1 centers[32]. Assuming the measured time constant $\tau_{dip} \approx 73$ μs stems from inter-$N^0$ dipolar couplings, we derive[31,33] an approximate nitrogen concentration of 0.25 ppm, consistent with the manufacturers specifications; this concentration corresponds to an average nitrogen separation of 28 nm (Fig. 1f). For future reference, extensive observations within other frequency windows (corresponding, e.g., to the $VH^-$ center[21], not shown) yielded negative results, suggesting that, if at all present, other paramagnetic defects with spin number greater than 1/2 have a concentration lower than nitrogen. We emphasize additional work will be required to validate this notion, as it is nearly impossible to exclude all alternatives (e.g., large crystal fields could shift transitions beyond our experimental range).

A closely related piece of information is presented in Fig. 2. Here we modify the DEER protocol to introduce a variable wait time $\tau_w$ after optical charge initialization of the NVs but before application of the MW sequence; we also use a shorter (3 μs) green laser pulse immediately prior to the first MW1 pulse for $NV^-$ spin repolarization after the wait time (Fig. 2a). Plotting the results as a function of $\tau_w$ with all other DEER parameters unchanged, we find a gradual decay with a characteristic time $\tau_{tun} \approx 2.4$ ms (Fig. 3b). This result indicates that the N$^0$ concentration decreases after green initialization, a behavior suggestive of electron tunneling to neighboring traps. Note that we can rule out NV$^-$ relaxation as a factor in the observed response, not only because NV$^-$ spins relax slowly (the measured NV$^-$ $T_1$-time is ~6 ms, longer than $\tau_{tun}$), but, most importantly, because the laser pulse preceding the DEER sequence is sufficiently long to re-pump all NVs into a spin polarized state.

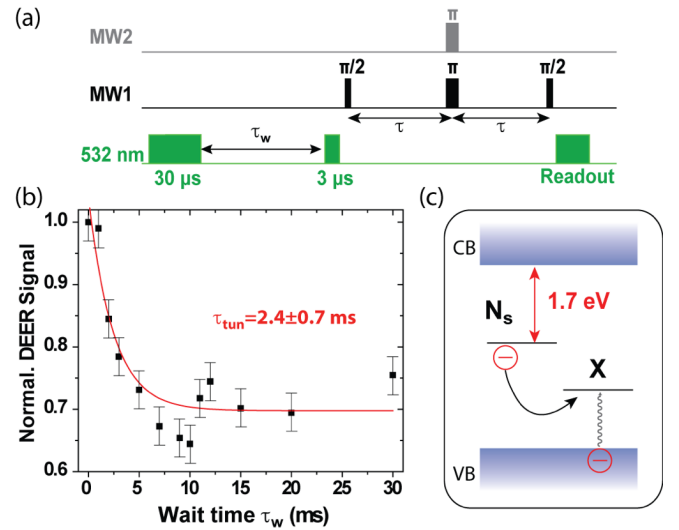

**Figure 2.** (a) Adapted DEER sequence for monitoring the N charge state. (b) Normalized DEER contrast as a function of the wait time $\tau_w$. The amplitudes are recorded at the $m'_I = 0$ transition; all other conditions as in Fig. 2. (c) Schematic energy level diagram. The X acceptor level can be populated either by electron tunneling from neighboring $N^0$ of by optical injection of a valence electron.

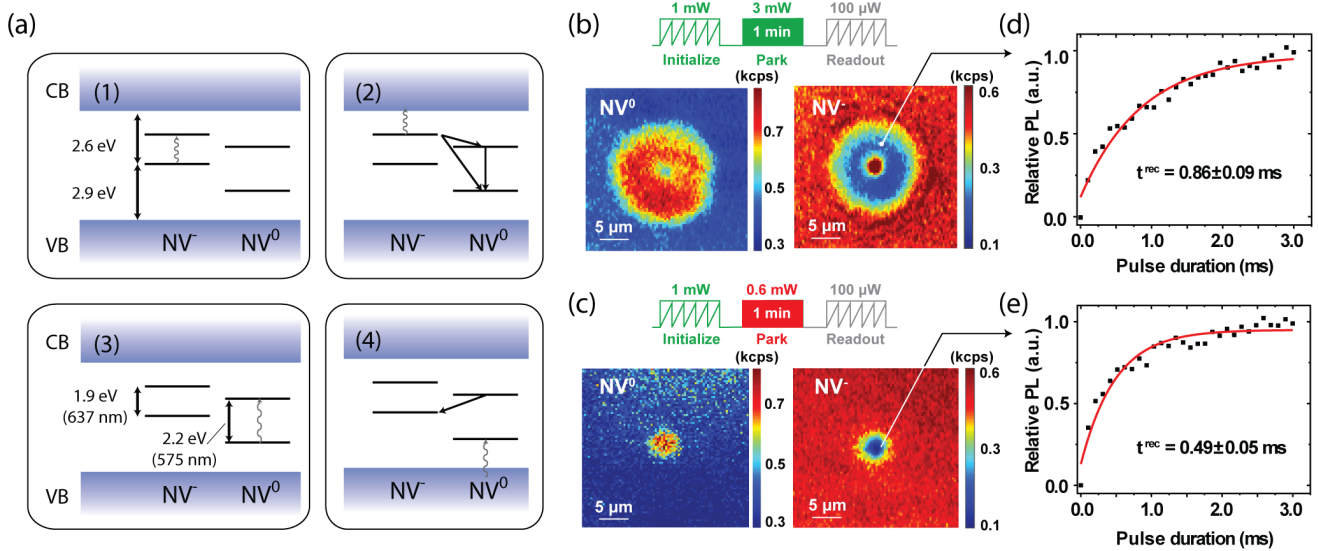

**Figure 3.** (a) Schematics of NV photochromism under optical (~460-637 nm) excitation. Steps (1) and (2) describe NV$^-$ photo-ionization into NV$^0$; steps (3) and (4) correspond to recombination of NV$^0$ into NV$^-$. CB (VB) stands for conduction (valence) band. (b) NV$^0$ and NV$^-$ selective confocal scans (respectively, left and right) upon application of the upper pulse protocol. Zig-zags (solid rectangles) indicate a beam scan (park); green (red) color indicates laser excitation at 532 nm (632 nm). We use a green (red) laser light for NV$^0$ (NV$^-$) readout. (c) Same as in (b) but for a red laser park (0.6 mW). In (b) and (c) the park time is 1 min and the scans correspond to 100×100 pixels and a 10 ms dwell time over a 32×32 µm$^2$ area. (d) NV$^-$ photoluminescence as a function of green laser pulse duration at a point within the NV$^-$-depleted halo; the laser power is 500 µW. (e) Same as in (d) but at the NV$^-$-depleted site created via the protocol in (c).

Given the significant fractional change of the observed signal (likely limited by unwanted N$^0$ reactivation via the 3 µs laser pulse), we conclude these electron traps are at least as abundant as nitrogen. We tentatively rationalize these results by invoking the presence of a point defect — here named 'X' — presumably a deep acceptor (Fig. 2c, see below). We hypothesize electron capture takes place via tunneling from N$^0$ (though the non-monotonic decay in Fig. 2b could be indicative of more complex dynamics, a question we defer to future studies).

The model in Fig. 2c also suggest an alternative channel of electron capture via photo-excitation from the valence band, a process necessarily accompanied by the generation of holes. To gather additional information on these traps, we implement the NV photo-conversion experiments in Fig. 3. In the visible range, NV photochromism is generally made possible via two-step, one-photon processes whose exact nature depends on the starting charge state and excitation wavelength[34]. For example, 'green' laser excitation (532 nm) can ionize negatively charged NVs via the sequential absorption of two photons, successively promoting the excess electron into the NV$^-$ excited state, then to the conduction band (Fig. 3a). Conversely, green light excitation of NV$^0$ followed by photo-injection of an electron from the valence band can lead to the formation of negatively charged NVs (a process normally referred to as 'recombination'). Continuous 532 nm illumination leads, therefore, to recurring NV charge state interconversion, a process that must be accompanied by a steady stream of diffusing electrons and holes from the NV site into the bulk. Importantly, recombination is largely suppressed under 632 nm illumination, because 'red' photons cannot excite NV$^0$ (whose zero-phonon line corresponds to 575 nm, Fig. 3a).

Fig. 3b shows the NV charge pattern emerging from a protocol that starts with a 1 mW green laser scan to (preferentially) initialize NVs into the negatively charged state, followed by a prolonged (1 min) green laser park at the midpoint of the scanned region. Comparison of the left and right panels (respectively selective to NV$^0$ and NV$^-$), shows that negatively charged NVs in the vicinity of the illumination point gradually transition to the neutral charge state via selective capture of photo-generated holes[26]. Note that to attain charge neutrality, photo-electrons must be captured by other point defects, arguably positively charged nitrogen[26]. For future reference, NV$^0$ can also form via direct NV$^-$ ionization during red illumination[34]. This is shown in Fig. 3c where we park a 632-nm beam on an NV$^-$-initialized plane; we find only local NV charge transformation into the neutral state, consistent with previous work[26,35].

A complementary indication of dynamics beyond NV and N charge interconversion is presented in Figs. 3d and 3e (see also the schematic model in Fig. 2c): Using a 532-nm laser pulse of variable duration, we monitor NV$^0$ recombination after parking the beam in NV$^-$-depleted areas produced either via hole capture or red-laser-induced ionization. Remarkably, we find that the observed recombination time is substantially longer for NVs initialized via hole capture, a result we find systematically, irrespective of the exact position within the doughnut-shaped area around the point of illumination (below referred to as the 'halo'). Taking into account the ensemble nature of our measurements (there are ~500 NV per µm$^3$ at a concentration of 2.5 ppb) we interpret this finding as an indication of greater hole recapture efficiency, in turn, a consequence of changes in the charge state of defect X, impacted differently depending on the charge initialization protocol.

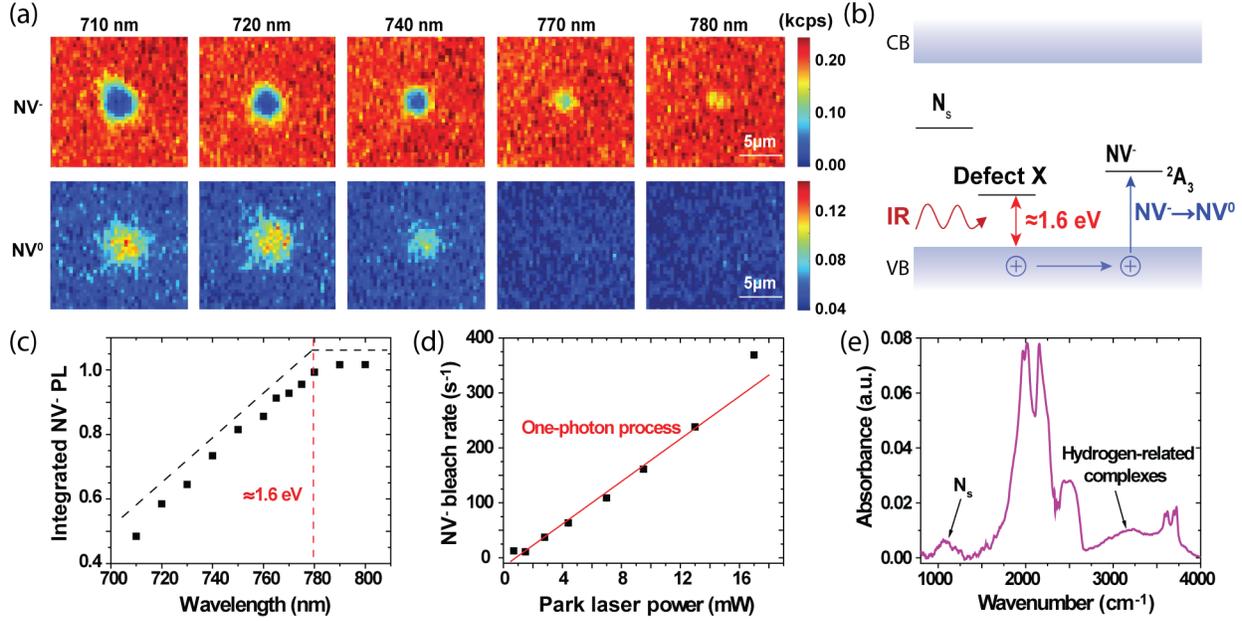

**Figure 4.** (a) We use a protocol identical to that of Fig. 1c but park an IR beam of variable wavelength. Top (bottom) panels show the NV$^-$ (NV$^0$) fluorescence after a 5 s, 20 mW laser park; all other conditions as in Fig. 1c. (b) Proposed mechanism of the NV$^-$ bleaching involving the promotion of the electron from the valence band into defect X and capture of the created hole by NV$^-$. (c) Integrated NV$^-$ photoluminescence in a 2.8×2.8 µm around the point of illumination as a function of the illumination wavelength. (d) Decay rate of the NV$^-$ fluorescence under 710 nm illumination as a function of the illumination power; the red line is a linear fit. (e) Fourier-transformed infrared absorbance spectrum.

Although a quantitative description is not presently possible, we gain a crude understanding by considering the rate equation governing the NV$^0$ density,

$$\frac{dQ_0}{dt} = -k_0 Q_0 + \kappa_p p(Q - Q_0),\qquad(1)$$

where $k_0$ is the recombination rate under green light, $\kappa_p$ is the NV$^-$ hole volume capture, $p$ is the hole density, and $Q = Q_0 + Q_-$ is the total NV density. For simplicity, Eq. (1) ignores back-conversion of NV$^-$ into NV$^0$ via photo-ionization as well as electron capture by NV$^0$ (a good approximation in these samples[35]). The second term in the sum amounts to NV$^0$ formation via NV$^-$ ionization from hole capture, a process that effectively slows down the recombination of NV$^0$, governed by the first term. The hole concentration, in turn, is largely dependent on the charge state of X (see Fig. 2c), an interplay that can be cast as

$$\frac{dp}{dt} = -Dp + k_X X_0 - \lambda_p p(X - X_0),\qquad(2)$$

and

$$\frac{dX_0}{dt} = -k_X X_0 + \lambda_p p(X - X_0),\qquad(3)$$

where $D$ represents an effective hole loss rate due to diffusion outside the point of illumination, $X_0$ denotes the local density of X defects in the neutral charge state, $X = X_0 + X_-$ is the total density of X defects, $k_X$ is the conversion rate of X$^0$ into the negatively charged state under green light, and $\lambda_p$ is the hole volume charge trapping rate by X$^-$. In Eq. (2) we ignore hole capture by neutral nitrogen (characterized by a low cross section[36,37]). Further, since presumably $X \gg Q$, we ignore the generation of holes via NV$^0$ recombination, as well as hole capture by NV$^-$, processes we assume dominated by the interconversion between X$^0$ and X$^-$.

From Eqs. (2) and (3), we conclude that the concentration of holes reaches a larger value if the starting charge state of X is neutral, as green optical excitation pumps electrons from the valence band in large numbers, i.e., the $k_X X_0$ term becomes dominant and the concentration of holes grows (see also Fig. 2c), effectively reducing the rate of NV$^0$ recombination in Eq. (1). This is precisely the starting condition in Fig. 3d, where holes (previously diffusing from the point of illumination under the park beam during preparation, Fig. 3b) arguably neutralize both NV and X in the halo region. In the converse limit (i.e., when most X defects are initially negatively charged), the number of holes — now limited to those emerging from $NV^0$ recombination, not considered here — decreases, while hole capture by $X^-$ is maximal (last term on the right-hand side of Eqs. (2) and (3)). This is the starting point in Fig. 3e, where the preceding $NV^-$ ionization is arguably accompanied by the conversion of X into the negatively charged state.

For completeness, we mention that detection of $NV^0$ fluorescence during green illumination (not shown for brevity), yields a monotonic exponential decay, indicative of a gradual conversion from neutral to negatively charged NVs. Combined with Figs. 3d and 3e, these observations allow us to rule out the formation of positively charged states[38] (potentially produced via the consecutive capture of two diffusing holes), as $NV^+$ would arguably reconvert first into neutral and thus lead to a transient growth of $NV^0$ during green illumination, in conflict with experiment.

To further these ideas, we implement one more time the laser protocol in Fig. 3c but we replace the 632-nm beam during the park by infra-red excitation in the range 710-800 nm (Fig. 4a). Remarkably, we find that NVs locally transform to neutral even though the photon energy is significantly below the minimum required (637 nm) to activate the two-step ionization process discussed above (Fig. 3a). Local $NV^- \rightarrow NV^0$ conversion at the site of illumination points, therefore, to the presence of holes, here generated through trapping of electrons photo-excited from the valence band into the X acceptor level (Fig. 4b). Interestingly, NVs act in this case as charge reporters exposing a process otherwise difficult to observe.

Longer excitation wavelengths gradually lead to less efficient $NV^-$ ionization, until virtually no NV charge conversion is observed at 780 nm. By plotting the integrated $NV^-$ fluorescence as a function of wavelength, we conclude the X acceptor level lies ~1.6 eV above the valence band (Fig. 4c). Further, at a given wavelength the $NV^-$ bleaching rate grows almost linearly with the laser power, hence supporting the notion of a single photon absorption process (Fig. 4d). Note this observation allows one to rule out other, unknown $NV^-$ ionization channels (involving, e.g., excitation into non-ground singlet states or via a phonon-assisted transition into the excited triplet) as they all necessarily fall in the category of two-step, one-photon processes, known to depend quadratically on the excitation laser power.

In summary, by exploiting the singular spin and charge properties of the NV, we reveal the presence of a coexisting point defect featuring a deep acceptor level. Though optically (and magnetically) dark, defect X is found to have an abundance at least comparable to substitutional nitrogen in CVD type 1b diamond. In the absence of site symmetry information, we can only hypothesize about its microscopic structure. N-based complexes seem less likely as they presumably have a concentration considerably lower than substitutional nitrogen. Hydrogen, on the other hand, is known to be a common contaminant in CVD diamond and several H-based point defects have already been identified and characterized. The absorption band in the 2800–3400 cm$^{-1}$ range — associated to CH stretching modes — in the infra-red spectrum of Fig. 4e supports the presence of hydrogen in our samples. Interestingly, computational studies[18] show that the ground state of the $VH^-$ ($VH^0$) point defect lies 1.67 eV (1.4 eV) above the valence band. However, observations at the expected resonance frequency — the $VH^-$ has spin number $S' = 1$ and a crystal field of 2.7 GHz[20] — yielded no DEER signal, suggesting defect X is different. The same applies to the NVH defect, forming a paramagnetic, spin-1/2 complex in the negative and positive charge states[17,20,21]. Additional work will therefore be mandatory to reveal the microscopic structure of defect X, likely to impact the performance of devices based on CVD diamond.

We thank Vasilios Deligiannakis and Maria Tamargo for assistance with the infra-red measurements. We acknowledge support from the National Science Foundation through grant NSF-1914945, and from Research Corporation for Science Advancement through a FRED Award; we also acknowledge access to the facilities and research infrastructure of the NSF CREST IDEALS, grant number NSF-HRD-1547830.